\documentclass[12pt]{article}

\usepackage{epsf}
\usepackage[dvips]{graphicx}
\usepackage{amssymb}

\begin{document}

\begin{center}
{\bfseries MULTIPLICITY DISTRIBUTION OF SECONDARY HADRONS AT LHC
ENERGY AND TOTAL CROSS SECTIONS OF HADRON-HADRON INTERACTIONS}

\vskip 5mm

V.A. Abramovsky$^{\dag}$, N.V. Radchenko$^{\ddag}$

\vskip 5mm

{\small {\it Novgorod State University }
\\
$\dag$ {\it E-mail: Victor.Abramovsky@novsu.ru }
\\
$\ddag$ {\it E-mail: nvrad@mail.ru }}
\end{center}

\vskip 5mm

\begin{center}
\begin{minipage}{150mm}
\centerline{\bf Abstract} The multiple production processes of
secondary hadrons in proton-antiproton scattering are divided into
three types. The first type is a shower of secondary hadrons
produced from gluon string decay, the second type is a shower of
secondary hadrons produced from two quark strings decay and the
third is a shower produced from three quark strings decay. At the
same time there are only two types for proton-proton scattering --
shower from gluon string and shower from two quark strings. These
showers do not correspond to pomeron showers originating from cuts
of one, two, three, \ldots pomerons. Multiplicity distribution in
gluon string is Gaussian, in two and three quark strings it is
negative binomial. Gluon string weight in the multiplicity
distribution is determined by the constant contribution to total
cross sections, the quark strings weights -- by the growing with
energy contributions. The expected value of proton-proton
scattering total cross section and the multiplicity distribution
at energy 14~TeV are given.\end{minipage}
\end{center}

\vskip 10mm

\section{Introduction}
In most QCD models hadrons interaction  is described by gluon
ladders with large number of bars which is increasing with total
energy $\sqrt{s}$~\cite{bib1}. These gluon ladders are associated
with pomerons. For inelastic processes these amplitudes correspond
to one or several parton cascades with large number of gluons.
Mean multiplicity of secondary hadrons $\langle n(s) \rangle$ is
proportional to mean number of partons $\langle \nu(s) \rangle$
with proportion coefficient which is of order one. Both these
values are large and they sufficiently quickly grow with growth of
total energy. It is very difficult in such models to adjust the
large value of mean multiplicity with such observable values: 1)
relations $\sigma_{tot}^{\pi^\pm p}/\sigma_{tot}^{p p}\simeq2/3$
and $\sigma_{tot}^{K^\pm p}/\sigma_{tot}^{p p}<2/3$ up to $\ln
s\leqslant 7$; 2) limitation of secondary hadrons transverse
momenta; 3) low value of Pomeranchuk trajectory slope.

In order to explain these contradictions there was proposed Low
Constituents Number Model (LCNM)~\cite{bib2},~\cite{bib3}  in
which there are only valent quarks and low number of
bremsstrahlung gluons in hadrons in initial state. The number of
bremsstrahlung gluons grows slowly with growth of $\sqrt{s}$, and
the interaction occurs as result of color exchange between quark
and gluon. Secondary hadrons are produced when strings (tubes) of
color electric field decay.

In this model we shall consider behavior of total cross sections
and multiplicity distributions in processes of $pp$ and $p\bar{p}$
interactions at high energies. These observables will be predicted
for energy $\sqrt{s}=14$~TeV.

\section{Three types of hadrons production processes in proton-proton and proton-antiproton scattering}
We describe the main processes of hadrons production in LCNM for
$pp$ and $p\bar{p}$ scattering  by diagrams shown in Fig.~1 --
Fig.~3.

In Fig.~1a  $p\bar{p}$ interaction process is described when there
are only valent quarks in initial state. Gluon exchange occurs
between these colorless states, and components, which had obtained
color charge, fly apart and form color field string. Since octet
states fly apart it will be gluon string. Corresponding inelastic
process of string decay is shown in Fig.~2. Dotted lines separate
interaction in final state, which leads to gluons production
(waved lines). Then these gluons form observable hadrons.

\begin{figure}[h]
\centerline{
\includegraphics[scale=1]{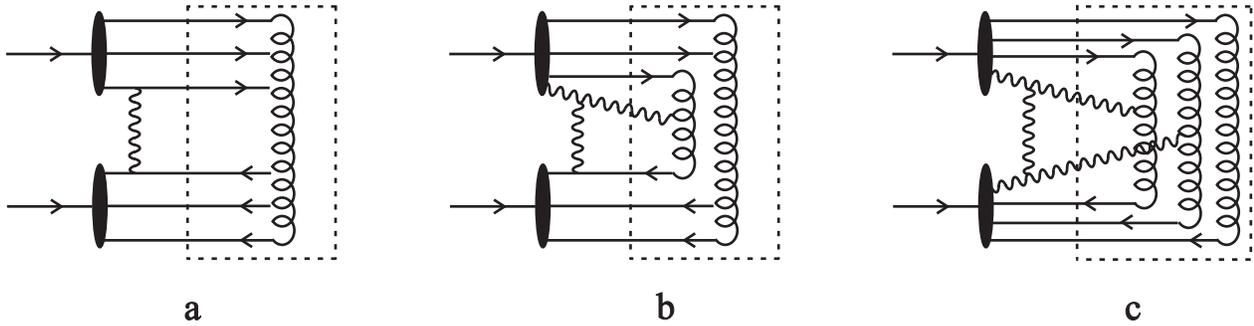}}
\caption{Three types of hadrons production processes in
proton-antiproton scattering: a) hadrons production in gluon
string; b) hadrons production in two quark strings; c) hadrons
production in three quark strings. Interaction in final state, in
which flying strings pass into hadrons, is separated by dotted
lines. Interaction is achieved by gluon exchange (waved line)
between different hadron components. Interacted gluons recharge
quark strings in b and c variants.}
\end{figure}

This diagram gives the certain value of secondary hadrons
multiplicity. Since there is a large number (practically infinite)
of alike diagrams and since they all have approximately the same
order, than the random value -- secondary hadrons multiplicity --
must obey normal distribution because of the probability theory
central limit theorem. Thus we suppose that multiplicity
distribution in the gluon string is Gaussian distribution.

\begin{figure}[h]
\centerline{
\includegraphics[scale=1]{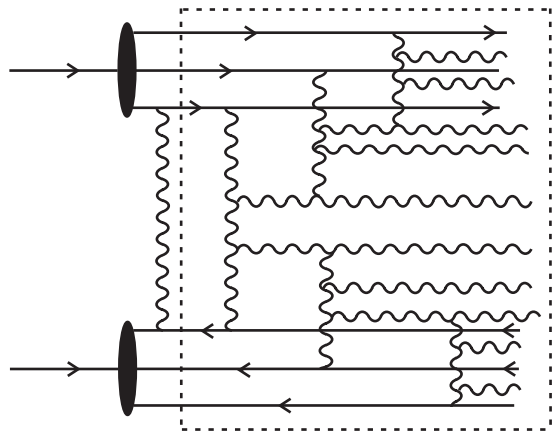}}
\caption{Diagram specifying gluon string decay from Fig.~1a.}
\end{figure}

We suppose that processes corresponding to diagrams with one
additional bremsstrahlung gluon (Fig.~1b) are inelastic processes
with production of two divided quark strings. Generally speaking,
transverse sizes of quark string must be in order of confinement
radius, i.~e. in order of hadron size. But since transverse
momentum of additional bremsstrahlung gluon is rather large,
approximately 1.5 -- 2 GeV, than compton wave length of the gluon
is small. This gluon must be absorbed at hadronization of one of
the quark strings. Therefore corresponding quark strings must have
transverse sizes comparable with gluon compton length.

Two bremsstrahlung gluons lead both to configuration with two
quark strings and to configuration with three quark strings shown
in Fig.~1c. Weights of configuration with two quark strings and
three quark strings do not depend on energy.

Hadrons production processes in $pp$ interaction differ from
production processes in $p\bar{p}$ interaction (Fig.~3). There is
no configuration with three quark strings in $pp$ interaction
since strings are produced between quark and diquark.

\begin{figure}[h]
\centerline{
\includegraphics[scale=1]{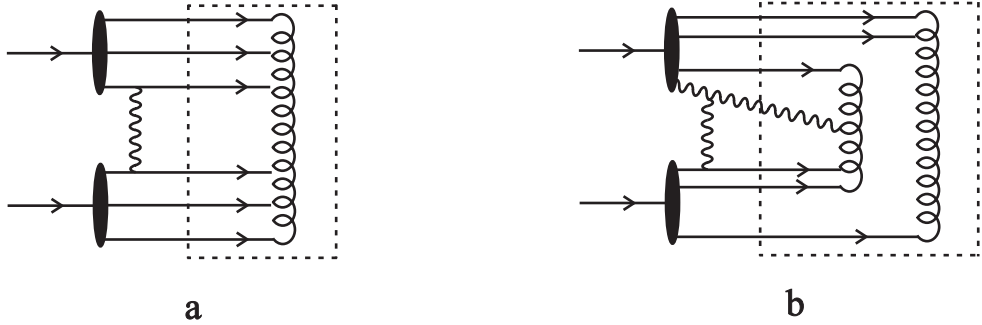}}
\caption{Two types of hadrons production processes in
proton-proton scattering: a) hadrons production in gluon string;
b) hadrons production in two quark strings.}
\end{figure}

We suppose that charged hadrons multiplicity distribution in quark
string fulfills negative binomial distribution
\begin{equation}\label{1}
P_n(k)=\frac{k(k+1)\ldots(k+n-1)}{n!}\left(\frac{\langle n\rangle
}{\langle n\rangle +k}\right)^n\left(\frac{k}{\langle n\rangle
+k}\right)^k.
\end{equation}
This distribution has two parameters: shape parameter $k$ and
mathematical expectation $\langle n\rangle $ -- mean multiplicity.
It is easy to show that convolution of two negative binomial
distributions with the same $\langle n\rangle $ and $k$ and
convolution of three negative binomial distributions with the same
$\langle n\rangle $ and $k$ are also negative binomial
distributions with $\langle n\rangle _2=2\langle n\rangle $,
$k_2=2 k$ and $\langle n\rangle _3=3\langle n\rangle $, $k_3=3 k$.
\begin{equation}\label{2}\begin{array}{l}
\displaystyle
P_n(k_2)=\!\!\!\!\!\sum_{\begin{array}{c}\scriptstyle
n_1,\;n_2\\[-1mm]\scriptstyle
n_1+n_2=n\end{array}}\!\!\!\!\!P_{n_1}(k)P_{n_2}(k)=\\
\displaystyle=\frac{k_2(k_2+1)\ldots(k_2+n-1)}{n!}\left(\frac{\langle
n\rangle _2}{\langle n\rangle _2+k_2}\right)^n\left(
\frac{k_2}{\langle n\rangle _2+k_2}\right)^{k_2},
\end{array}
\end{equation}
\begin{equation}\label{3}\begin{array}{l}
\displaystyle
P_n(k_3)=\!\!\!\!\!\sum_{\begin{array}{c}\scriptstyle
n_1,\;n_2,\;n_3\\[-1mm]\scriptstyle
n_1+n_2+n_3=n\end{array}}\!\!\!\!\!P_{n_1}(k)P_{n_2}(k)P_{n_3}(k)=\\
\displaystyle=\frac{k_3(k_3+1)\ldots(k_3+n-1)}{n!}\left(\frac{\langle
n\rangle _3}{\langle n\rangle _3+k_3}\right)^n\left(
\frac{k_3}{\langle n\rangle _3+k_3}\right)^{k_3}.
\end{array}
\end{equation}

\section{Behavior of total cross sections of proton-proton and proton-antiproton scattering}
Initial state components containing only valent quarks, containing
valent quarks and one or two additional gluons, lead to different
types of inelastic processes. It should be noted, that these
inelastic processes differ from inelastic processes occurring both
from cuts of pomeron and pomeron branchings. It also should be
noted, that the third additional gluon in initial state gives
negligible contribution. In accordance with this, total cross
sections of $pp$ and $p\bar{p}$ scattering can be written in form:
\begin{equation}\label{4}
\sigma_{tot}^{p(\bar{p})p}=63.52s^{0.358}\mp
35.43s^{0.56}+\sigma_{0}^{pp}+\sigma_1^{pp}\ln s+\sigma_2^{pp}(\ln
s)^2,
\end{equation}
where sign (-) stands for $pp$ and (+) stands for $p\bar{p}$
interaction. The first two terms describe contributions from non
vacuum reggeons exchange, the values and experimental data are
taken from~\cite{bib4}, $\sigma_{0}^{pp}$ is contribution of gluon
exchange between valent quarks, $\sigma_{1}^{pp}$ is contribution
of gluon exchange with components containing one additional gluon,
$\sigma_{2}^{pp}$ is contribution of gluon exchange with
components containing two additional gluons. Parameters
$\sigma_{0}^{pp}=20.08\pm0.42$, $\sigma_{1}^{pp}=1.14\pm0.13$,
$\sigma_{2}^{pp}=0.16\pm0.01$ were obtained from simultaneous
fitting of $pp$ and $p\bar{p}$ total cross sections. The result is
shown in Fig.~4. Thus we can predict the value of total cross
section $\sigma_{tot}^{pp}$ for energy $\sqrt{s}=14$~TeV
$\sigma_{tot}^{pp}= 101.30\pm6.65$~mb.

\begin{figure}[h]
\centerline{
\includegraphics[scale=0.65]{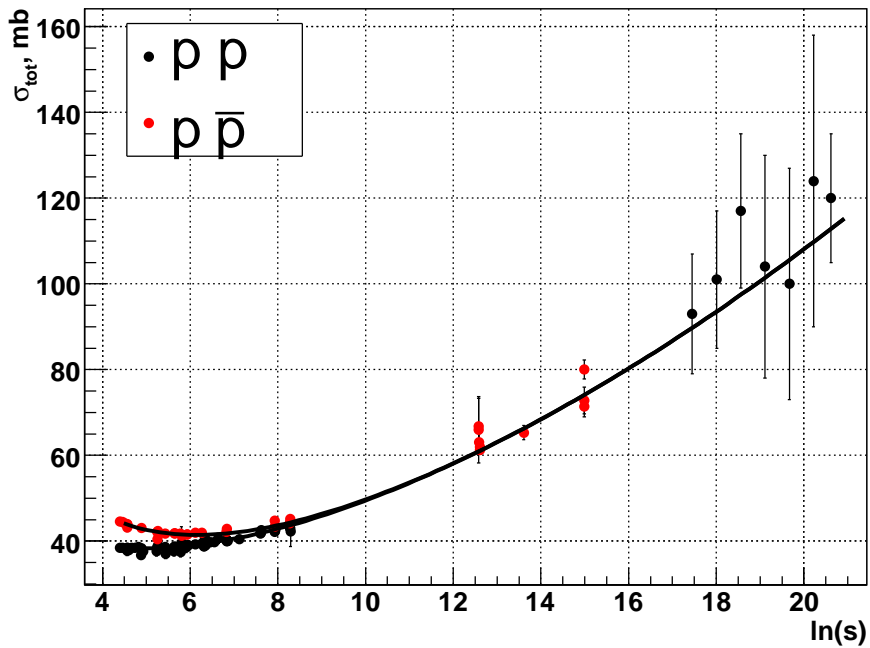}}
\caption{Total cross sections of proton-proton and
proton-antiproton scattering.}
\end{figure}

We want to stress once more, that there is the possibility of only
two additional bremsstrahlung gluons at energies up to LHC energy.
With this assumption we fitted non single diffraction cross
sections of $pp$ and $p\bar{p}$
scattering~\cite{bib5},~\cite{bib6},~\cite{bib7},~\cite{bib8},~\cite{bib9}.
$$
\sigma_{nsd}=\sigma_{tot}-\sigma_{el}-\sigma_{sd}.
$$
The contribution of non vacuum reggeons to elastic cross section
$\sigma_{el}$ and to single diffraction cross section
$\sigma_{sd}$ at energies higher than 44.5~GeV is low and we
neglect it. Non vacuum reggeons give contribution only to
$\sigma_{nsd}$. We have subtracted this contribution from the
experimental values of $\sigma_{nsd}$ and obtained values of
$\sigma_{nsd}^{vac}$ -- vacuum contributions to $\sigma_{nsd}$.
Then we have fitted these values with formulae
\begin{equation}\label{5}
\sigma_{nsd}^{vac}=\sigma_{0}^{nsd}(1+\delta_1^{nsd}\ln
s+\delta_2^{nsd}(\ln s)^2.
\end{equation}

The parameter values are $\sigma_{0}^{nsd}=14.50$ (fixed from
physical considerations), $\delta_1^{nsd}=0.0495\pm0.0105$,
$\delta_2^{nsd}=0.0073\pm0.0009$. The first term in~(\ref{5})
corresponds to gluon string contribution and defines area under
normal distribution curve. The third term corresponds to two
bremsstrahlung gluons contribution and, as it was explained above,
defines contribution of three quark strings configuration
$c_1\sigma_0^{nsd}\delta_2^{nsd}(\ln s)^2$ and of two quark
strings configuration $c_2\sigma_0^{nsd}\delta_2^{nsd}(\ln s)^2$,
at that, obviously, $c_1+c_2=1$. The value of
$c_1\sigma_0^{nsd}\delta_2^{nsd}(\ln s)^2$ gives area under
negative binomial distribution curve, corresponding to three quark
strings configuration. Area under negative binomial distribution
curve, corresponding to two quark strings configuration is given
by expression $\sigma_0^{nsd}(\delta_1^{nsd}\ln
s+c_2\sigma_0^{nsd}\delta_2^{nsd}(\ln s)^2)$. The coefficients
$c_1=0.23$ and $c_2=0.77$ were obtained from the fit and they do
not depend on the energy of colliding particles.

\section{Multiplicity distributions of charged hadrons of proton-proton and proton-antiproton scattering}
We have fitted data on charged multiplicity distributions for non
single diffraction processes in proton-antiproton scattering from
experiments UA5~\cite{bib6},~\cite{bib10} and E735~\cite{bib11}.
We suppose that multiplicity distribution is formed from the
normal distribution for events with gluon string and two negative
binomial distributions for events with two and three quark
strings. We also suppose that parameters for every quark string
both in configuration with two quark strings and in configuration
with three quark strings are the same.

We have also fitted data on charged  multiplicity distributions
for non single diffraction processes in proton-proton scattering
for energies $\sqrt{s}=44.5$, $\sqrt{s}=52.6$,
$\sqrt{s}=62.2$~GeV~\cite{bib9}. The results of fitting are shown
in Fig.~5 -- Fig.~24.

From these fits we have obtained parameters of normal distribution
for gluon string $\langle n\rangle_g$ and $\sigma$ and parameters
of negative binomial distribution for quark string $k$ and
$\langle n\rangle_q$ for energy range from $\sqrt{s}=44.5$ to
$\sqrt{s}=1800$~GeV. Then we have obtained the energy dependence
of quark and gluon strings parameters and thus we obtained the
values of parameters at energy $\sqrt{s}=14$~TeV. The multiplicity
distribution for LHC is shown in Fig.~25, 26. We can also
calculate the value of mean charged multiplicity for LHC, it is
$\langle n \rangle=71.57\pm4.37$.

\newpage
\begin{figure}[h]
\centerline{
\includegraphics[scale=0.65]{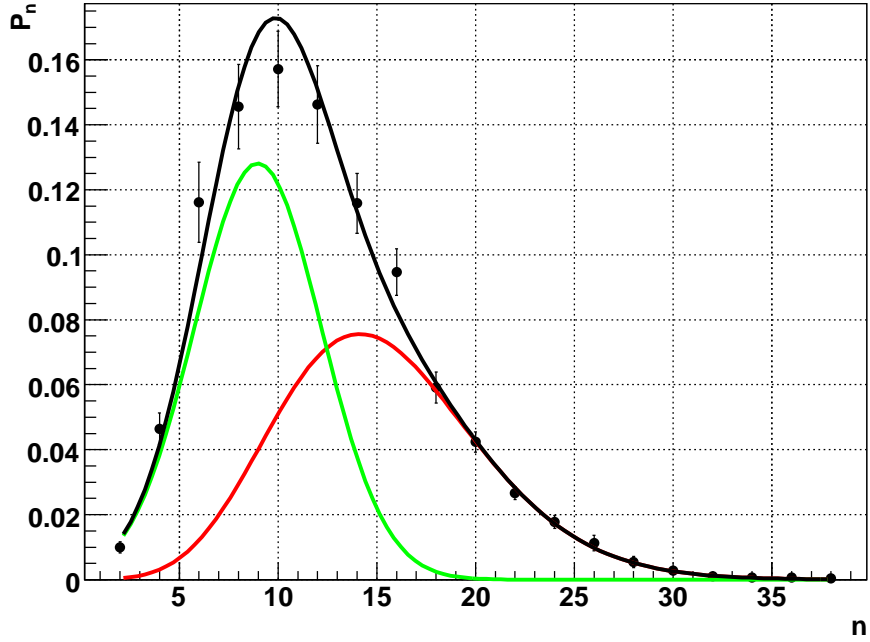}}
\caption{Charged  multiplicity distribution for proton-proton
scattering, $\sqrt{s}=44.5$~GeV~\cite{bib9}. Red line -- negative
binomial distribution for two quark strings, green line --
Gaussian distribution for gluon string, black line is sum of these
distributions, $\chi^2/ndf=13/14$. }
\end{figure}
\begin{figure}[!h]
\centerline{
\includegraphics[scale=0.65]{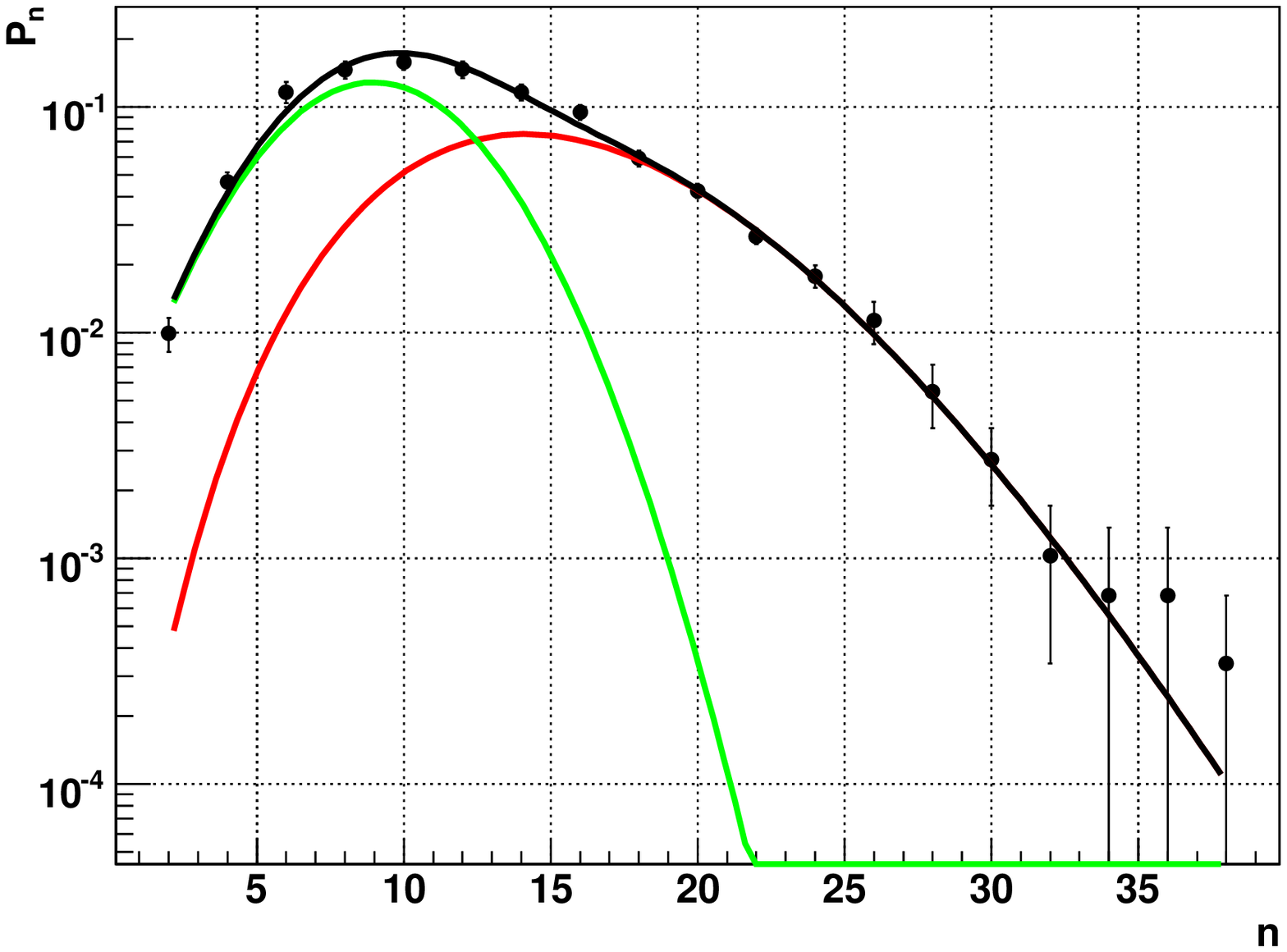}}
\caption{The same as for Fig. 5, logarithmic scale.}
\end{figure}

\newpage
\begin{figure}[h]
\centerline{
\includegraphics[scale=0.65]{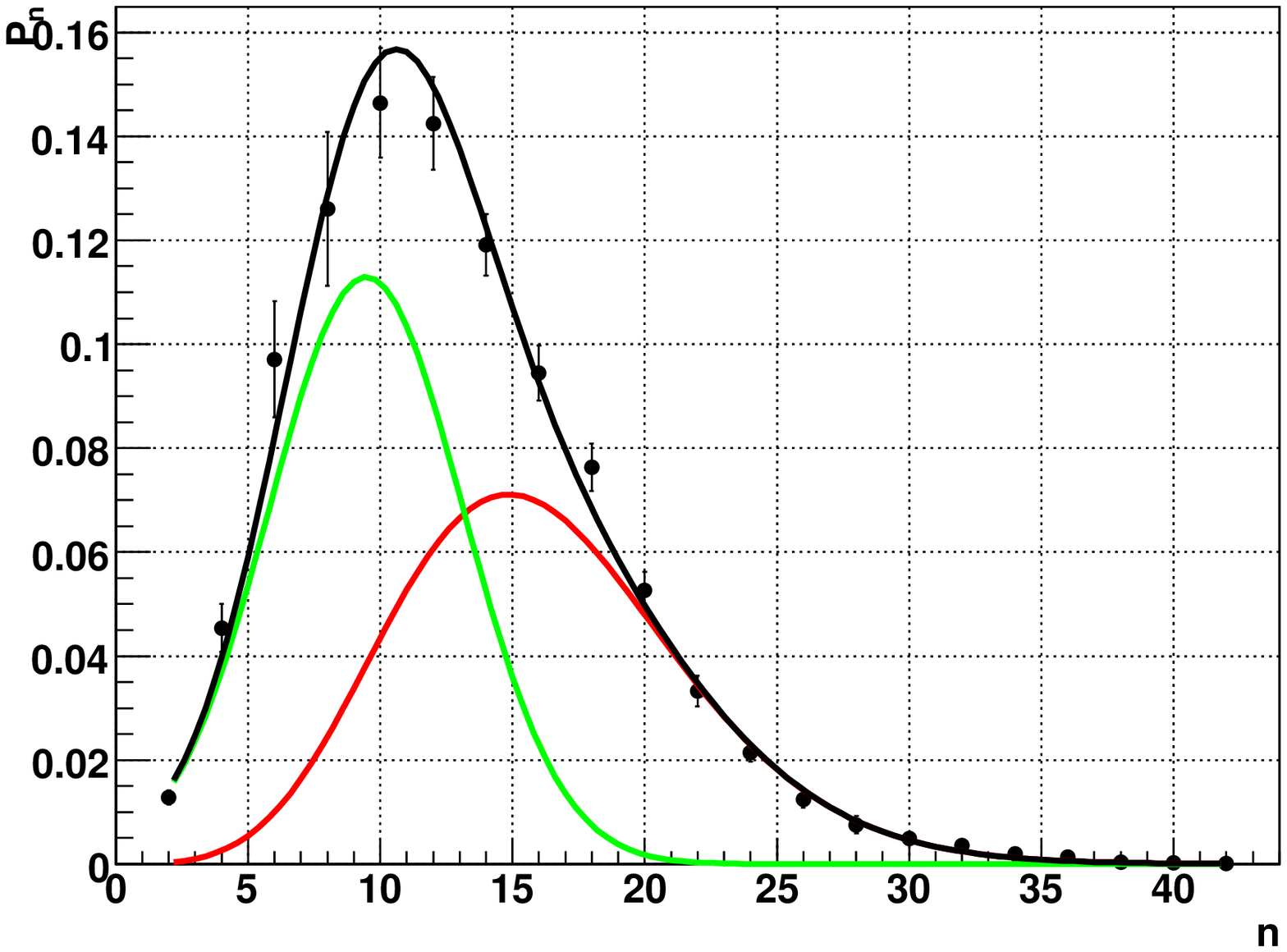}}
\caption{Charged  multiplicity distribution for proton-proton
scattering, $\sqrt{s}=52.6$~GeV~\cite{bib9}. Red line -- negative
binomial distribution for two quark strings, green line --
Gaussian distribution for gluon string, black line is sum of these
distributions, $\chi^2/ndf=17/16$. }
\end{figure}
\begin{figure}[!h]
\centerline{
\includegraphics[scale=0.65]{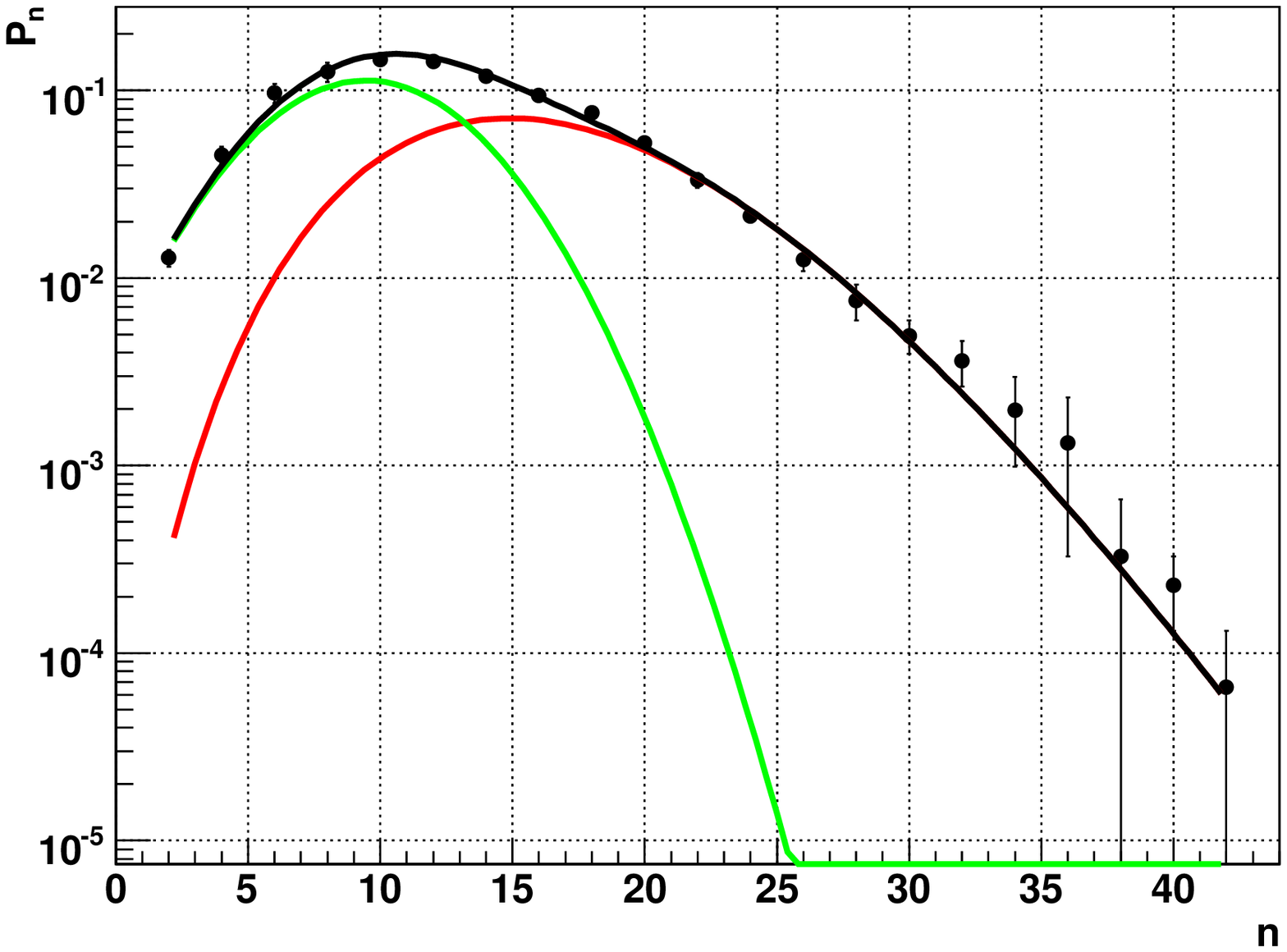}}
\caption{The same as for Fig. 7,  logarithmic scale.}
\end{figure}

\newpage
\begin{figure}[h]
\centerline{
\includegraphics[scale=0.65]{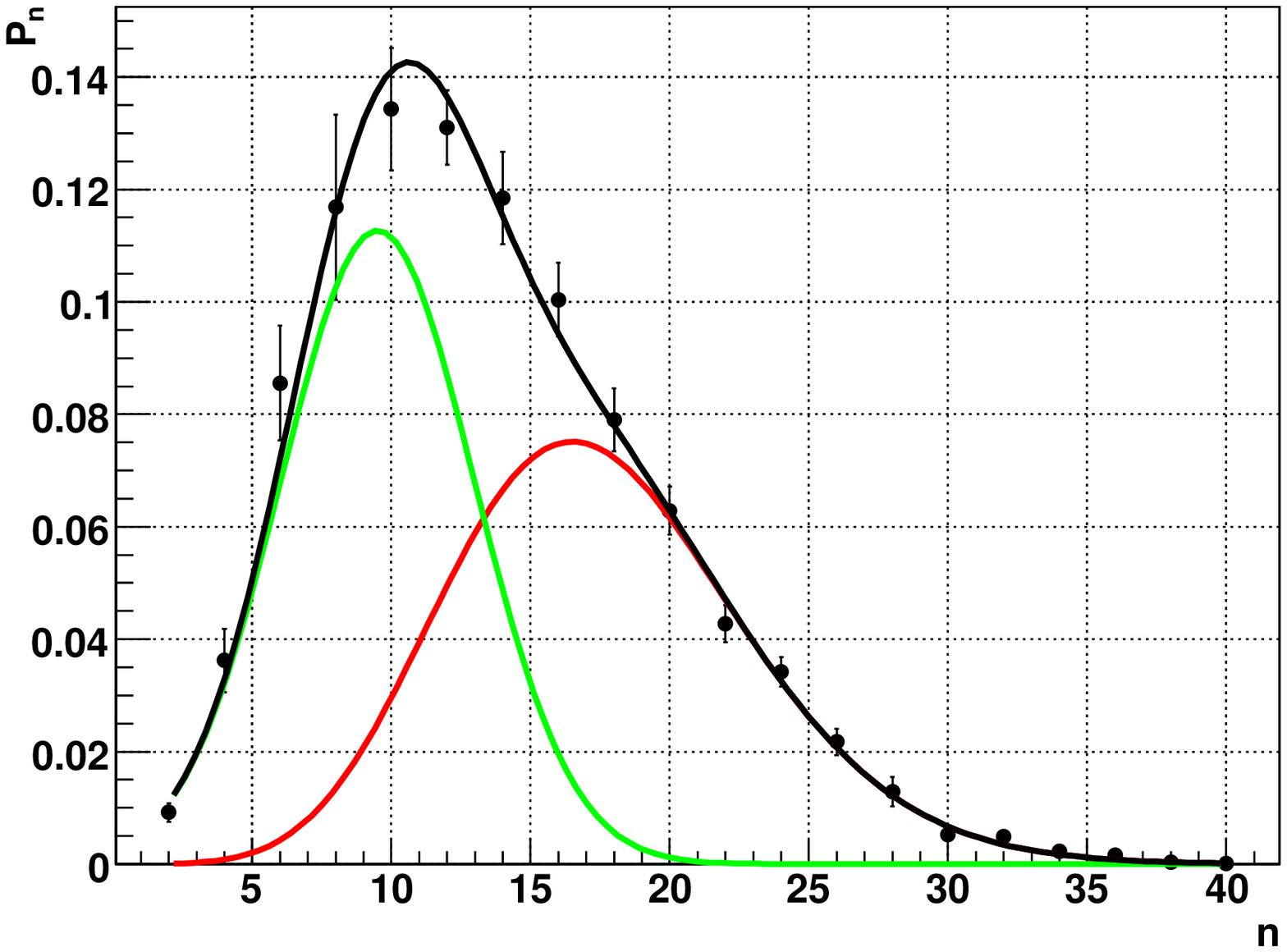}}
\caption{Charged  multiplicity distribution for proton-proton
scattering, $\sqrt{s}=62.2$~GeV~\cite{bib9}. Red line -- negative
binomial distribution for two quark strings, green line --
Gaussian distribution for gluon string, black line is sum of these
distributions, $\chi^2/ndf=13/15$. }
\end{figure}
\begin{figure}[!h]
\centerline{
\includegraphics[scale=0.65]{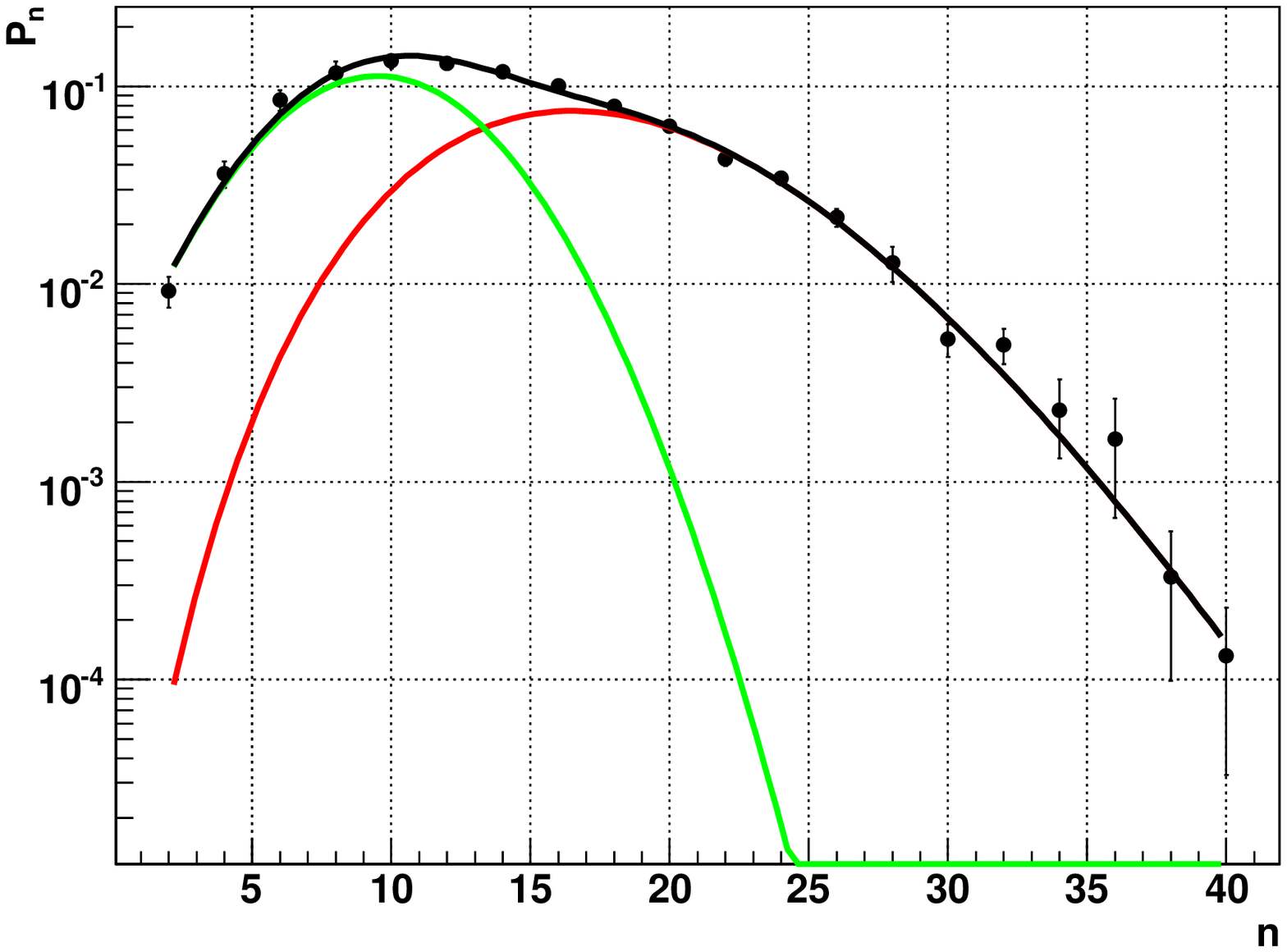}}
\caption{The same as for Fig. 9,  logarithmic scale.}
\end{figure}

\newpage
\begin{figure}[h]
\centerline{
\includegraphics[scale=0.65]{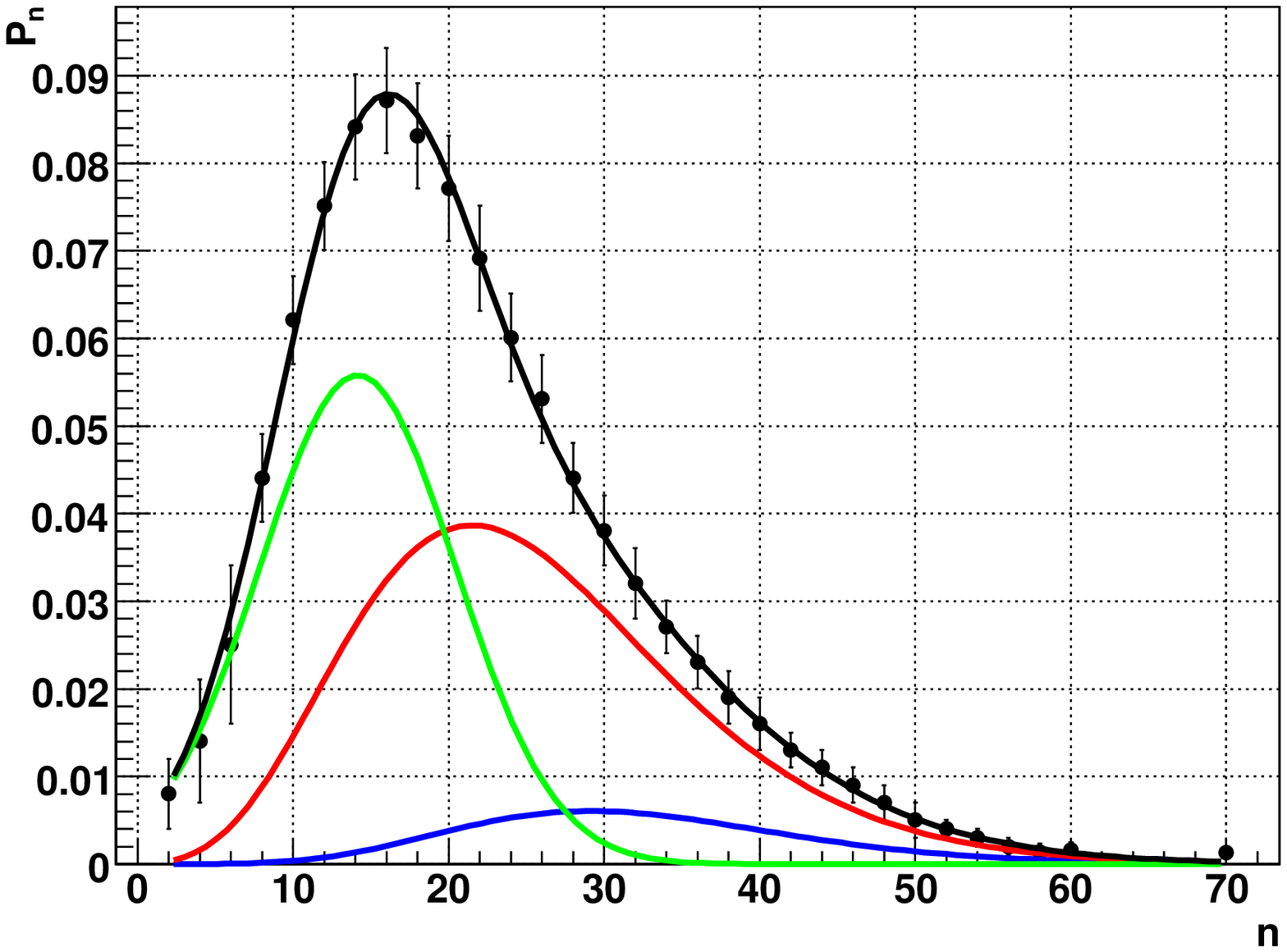}}
\caption{Charged  multiplicity distribution for proton-antiproton
scattering, $\sqrt{s}=200$~GeV~\cite{bib10}. Blue line -- negative
binomial distribution for three quark strings, red line --
negative binomial distribution for two quark strings, green line
-- Gaussian distribution for gluon string, black line is sum of
these distributions, $\chi^2/ndf=4/26$. }
\end{figure}
\begin{figure}[!h]
\centerline{
\includegraphics[scale=0.65]{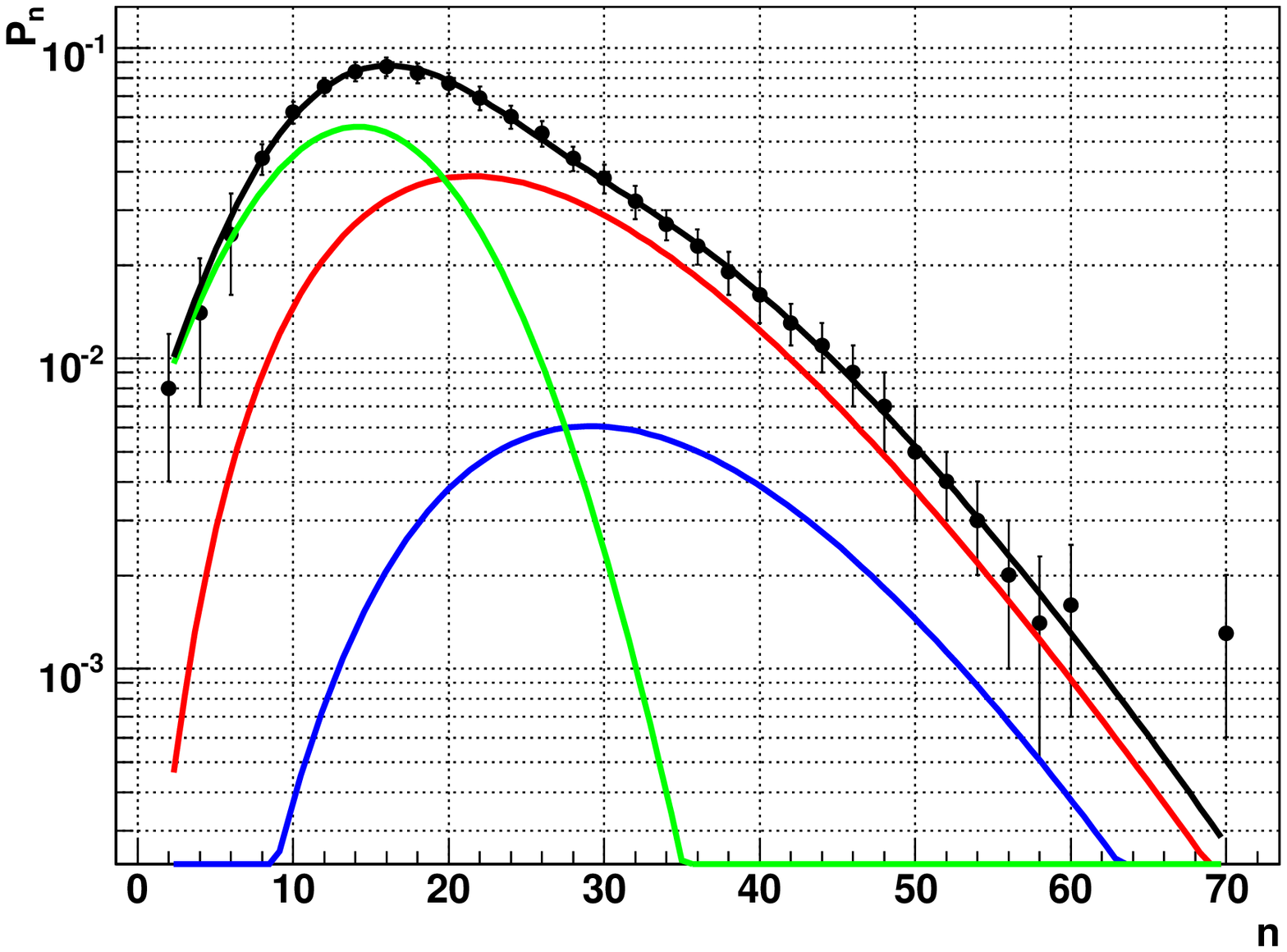}}
\caption{The same as for Fig. 11,  logarithmic scale.}
\end{figure}

\newpage
\begin{figure}[h]
\centerline{
\includegraphics[scale=0.65]{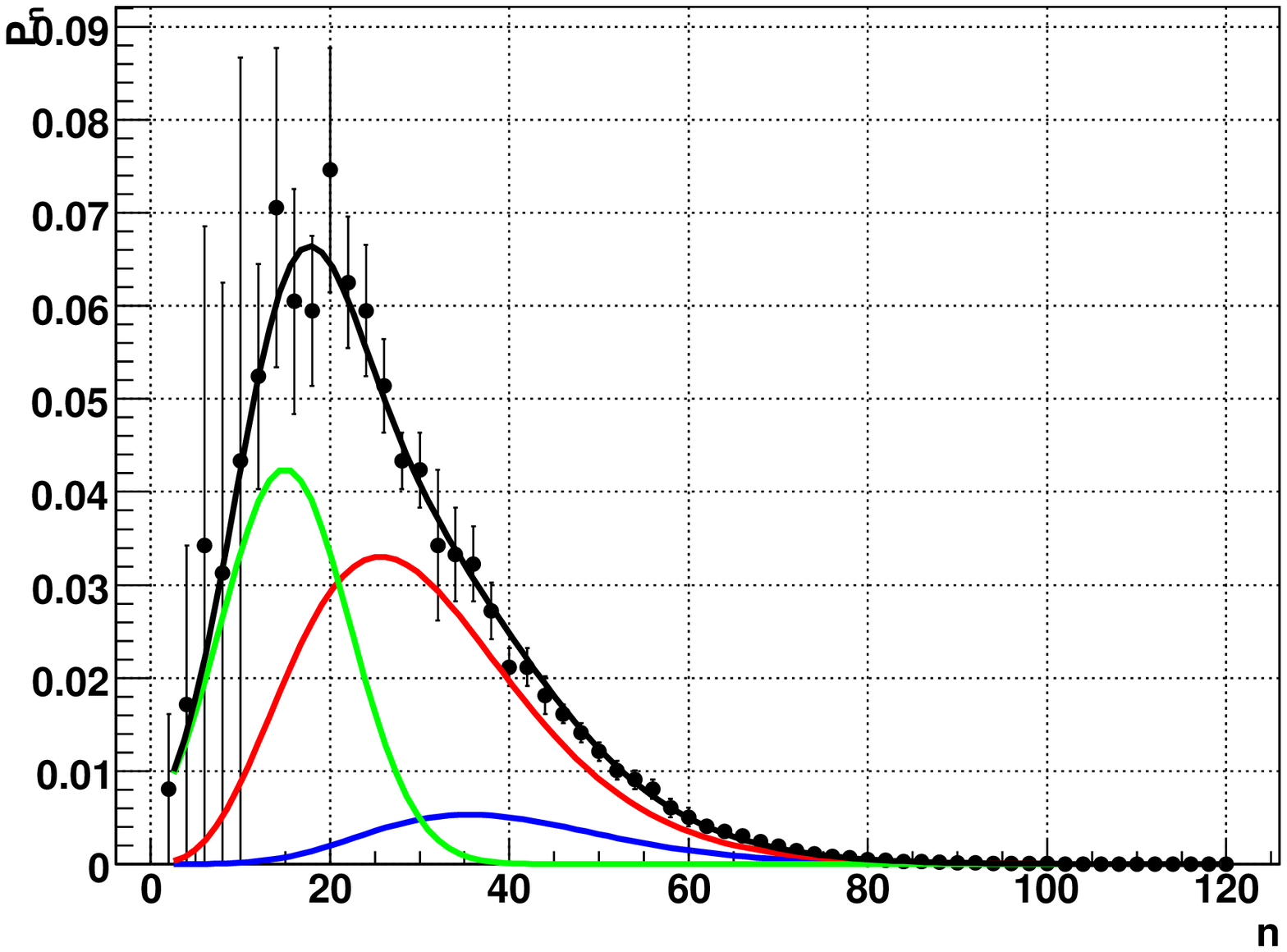}}
\caption{Charged  multiplicity distribution for proton-antiproton
scattering, $\sqrt{s}=300$~GeV~\cite{bib11}. Blue line -- negative
binomial distribution for three quark strings, red line --
negative binomial distribution for two quark strings, green line
-- Gaussian distribution for gluon string, black line is sum of
these distributions, $\chi^2/ndf=19/55$. }
\end{figure}
\begin{figure}[!h]
\centerline{
\includegraphics[scale=0.65]{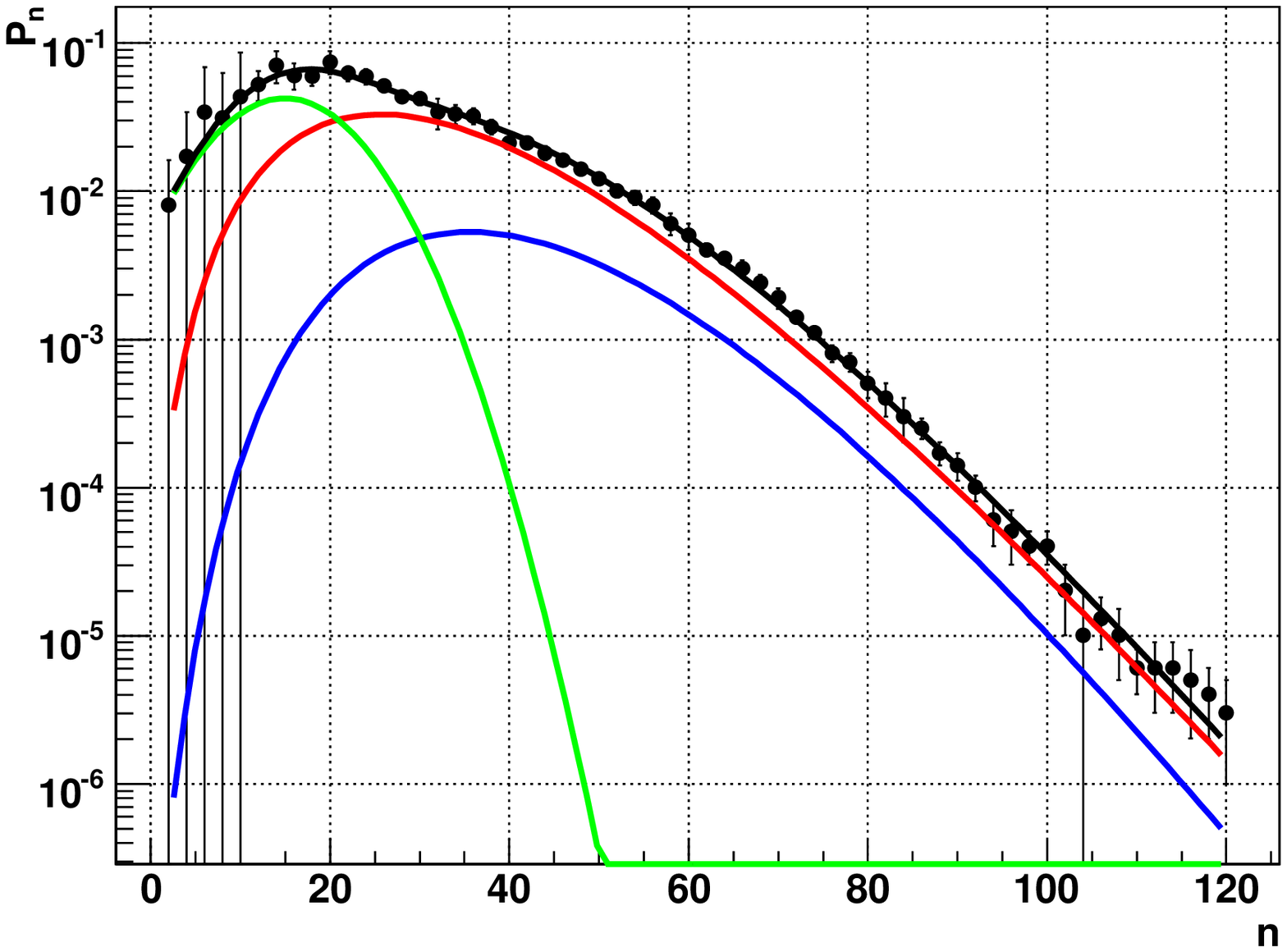}}
\caption{The same as for Fig. 13,  logarithmic scale.}
\end{figure}

\newpage
\begin{figure}[h]
\centerline{
\includegraphics[scale=0.65]{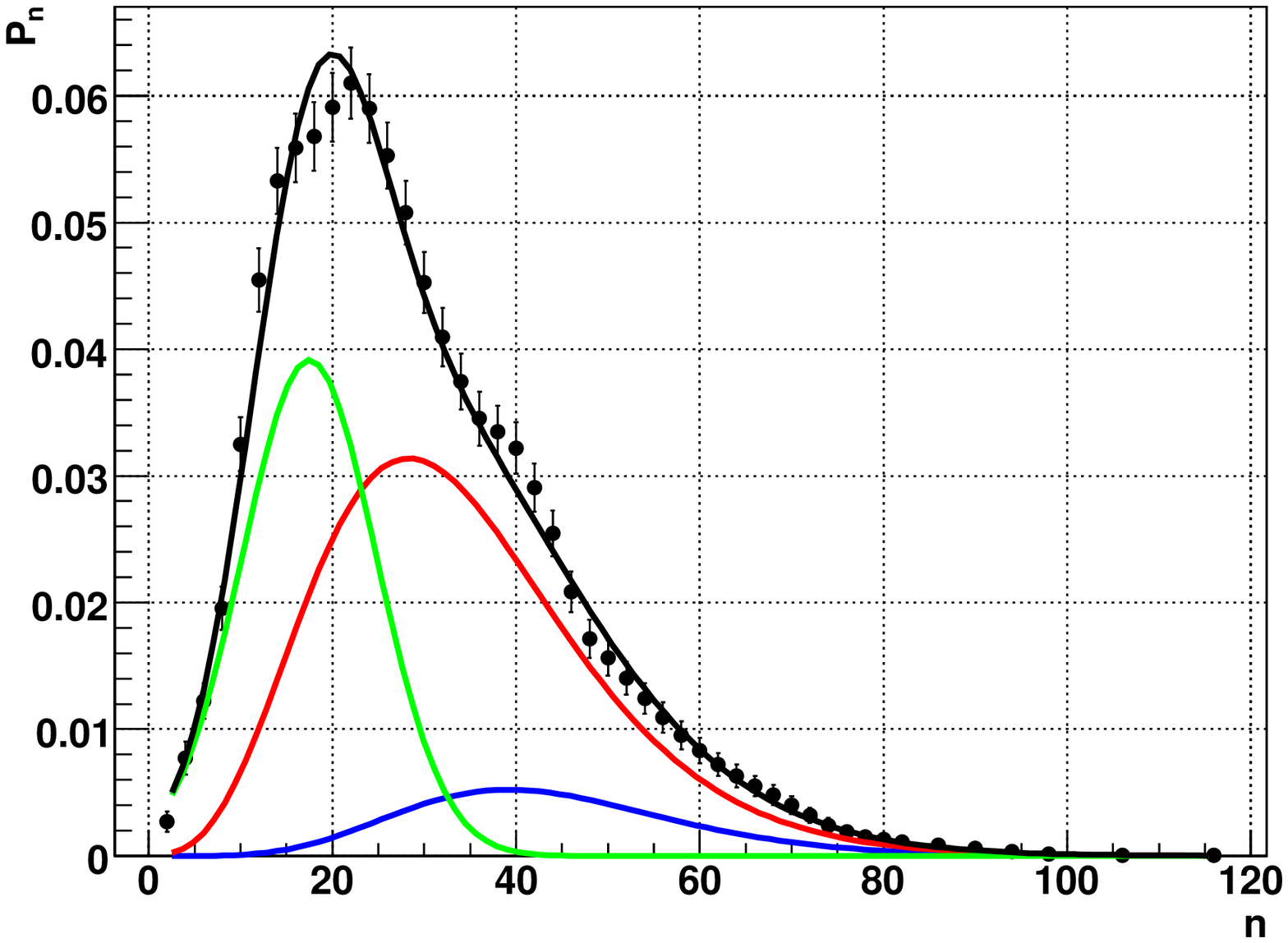}}
\caption{Charged  multiplicity distribution for proton-antiproton
scattering, $\sqrt{s}=546$~GeV~\cite{bib6}. Blue line -- negative
binomial distribution for three quark strings, red line --
negative binomial distribution for two quark strings, green line
-- Gaussian distribution for gluon string, black line is sum of
these distributions, $\chi^2/ndf=36/42$. }
\end{figure}
\begin{figure}[!h]
\centerline{
\includegraphics[scale=0.65]{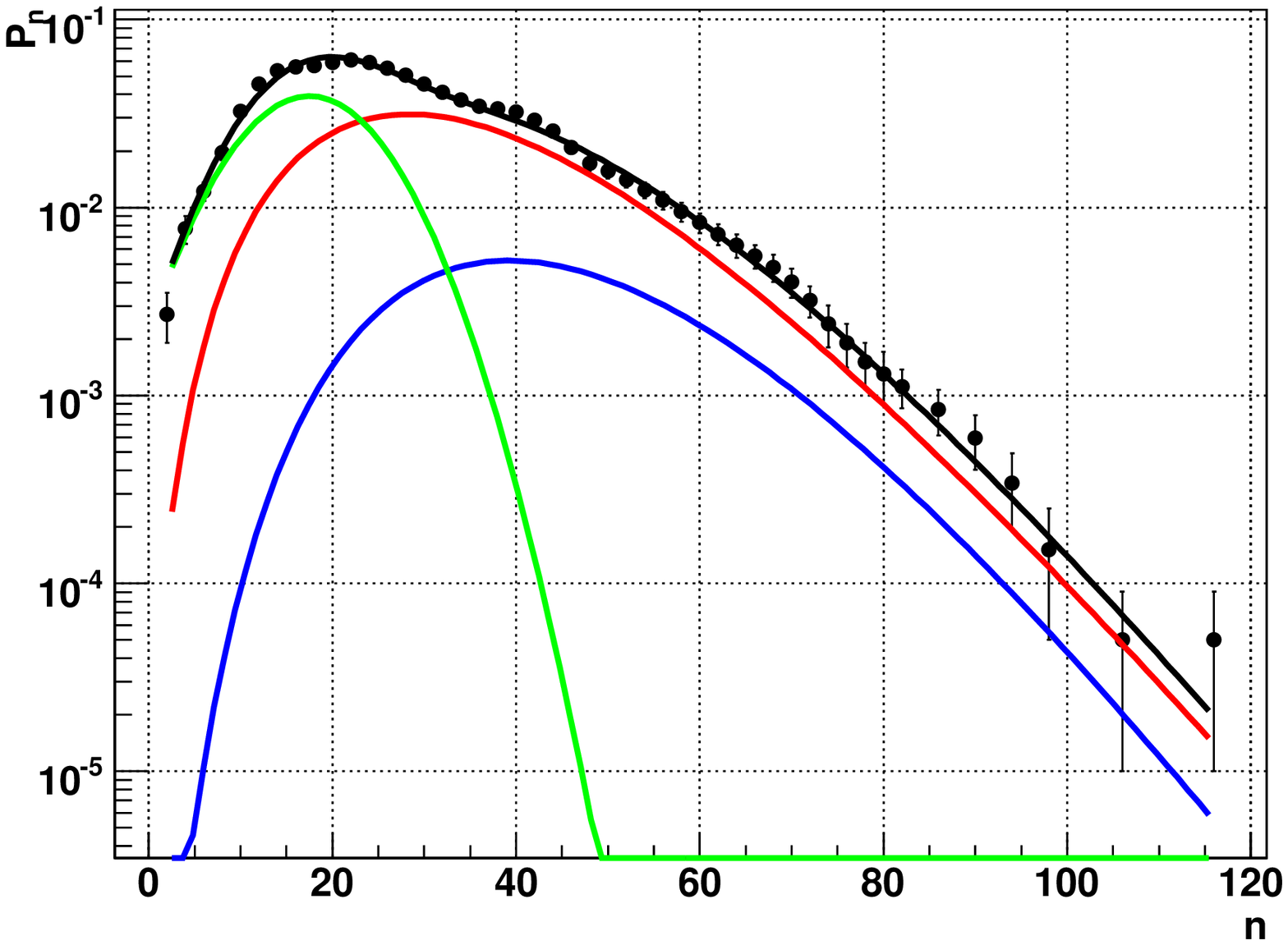}}
\caption{The same as for Fig. 15,  logarithmic scale.}
\end{figure}

\newpage
\begin{figure}[h]
\centerline{
\includegraphics[scale=0.65]{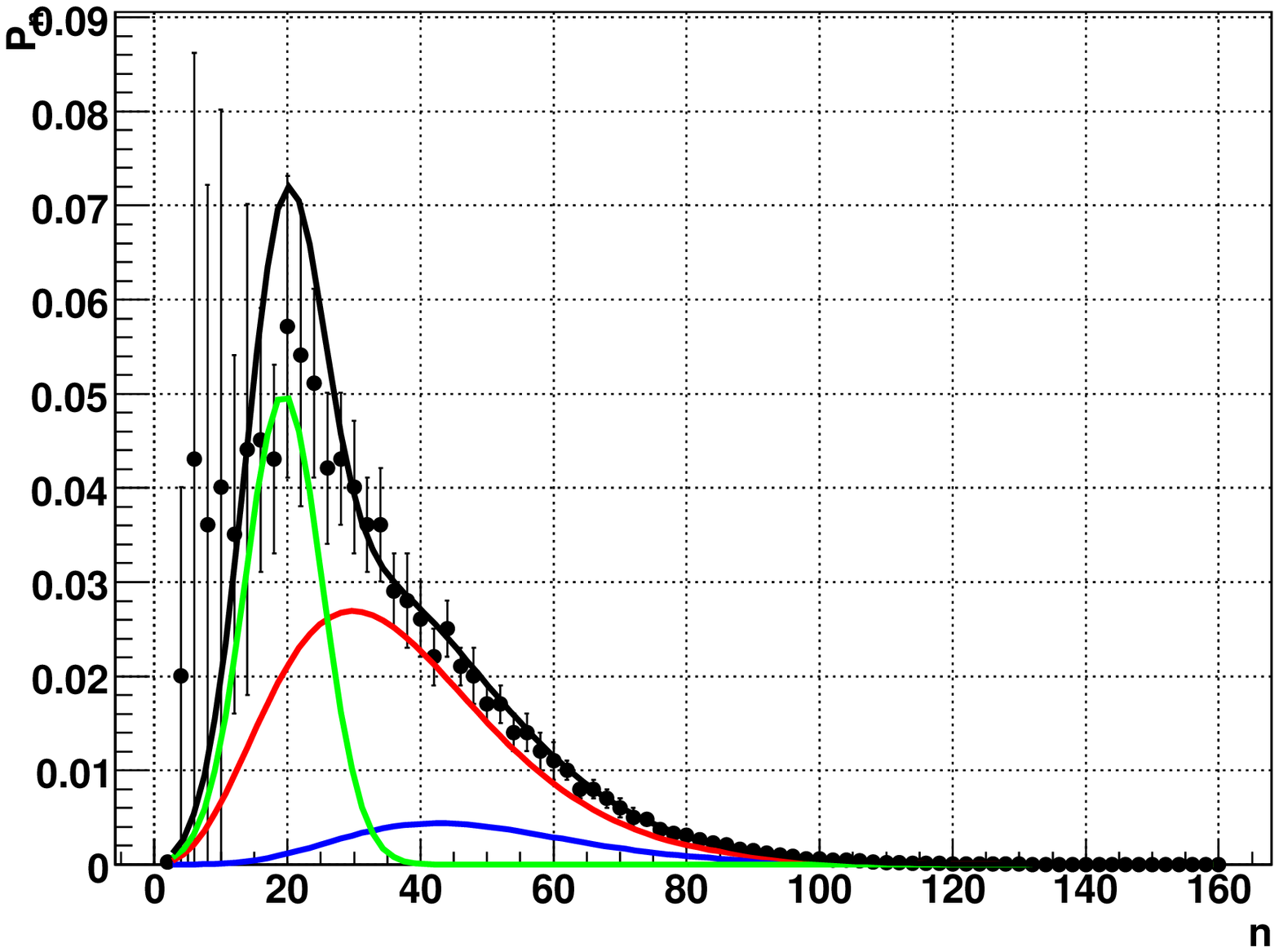}}
\caption{Charged  multiplicity distribution for proton-antiproton
scattering, $\sqrt{s}=546$~GeV~\cite{bib11}. Blue line -- negative
binomial distribution for three quark strings, red line --
negative binomial distribution for two quark strings, green line
-- Gaussian distribution for gluon string, black line is sum of
these distributions, $\chi^2/ndf=33/76$. }
\end{figure}
\begin{figure}[!h]
\centerline{
\includegraphics[scale=0.65]{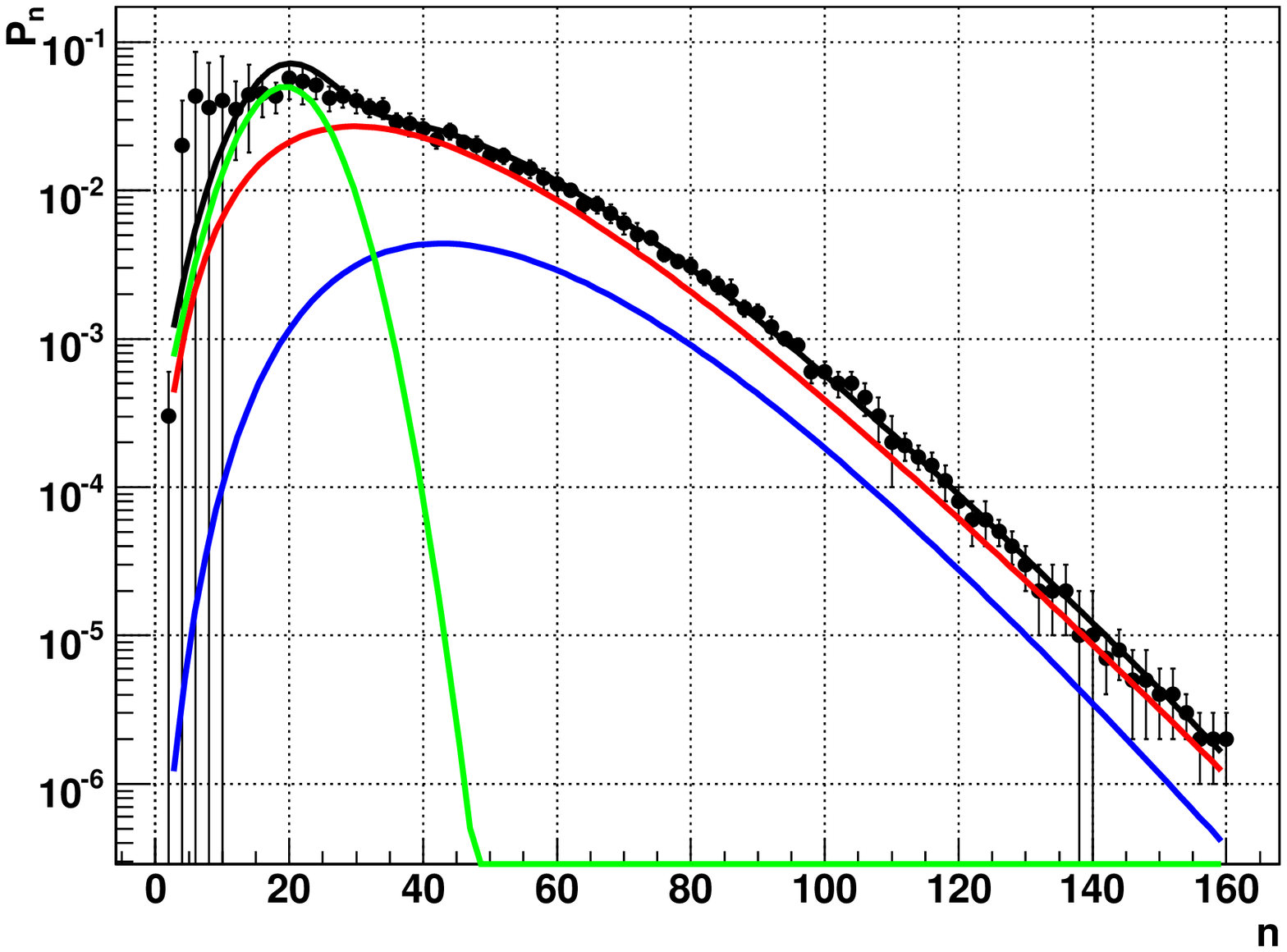}}
\caption{The same as for Fig. 17,  logarithmic scale.}
\end{figure}

\newpage
\begin{figure}[h]
\centerline{
\includegraphics[scale=0.65]{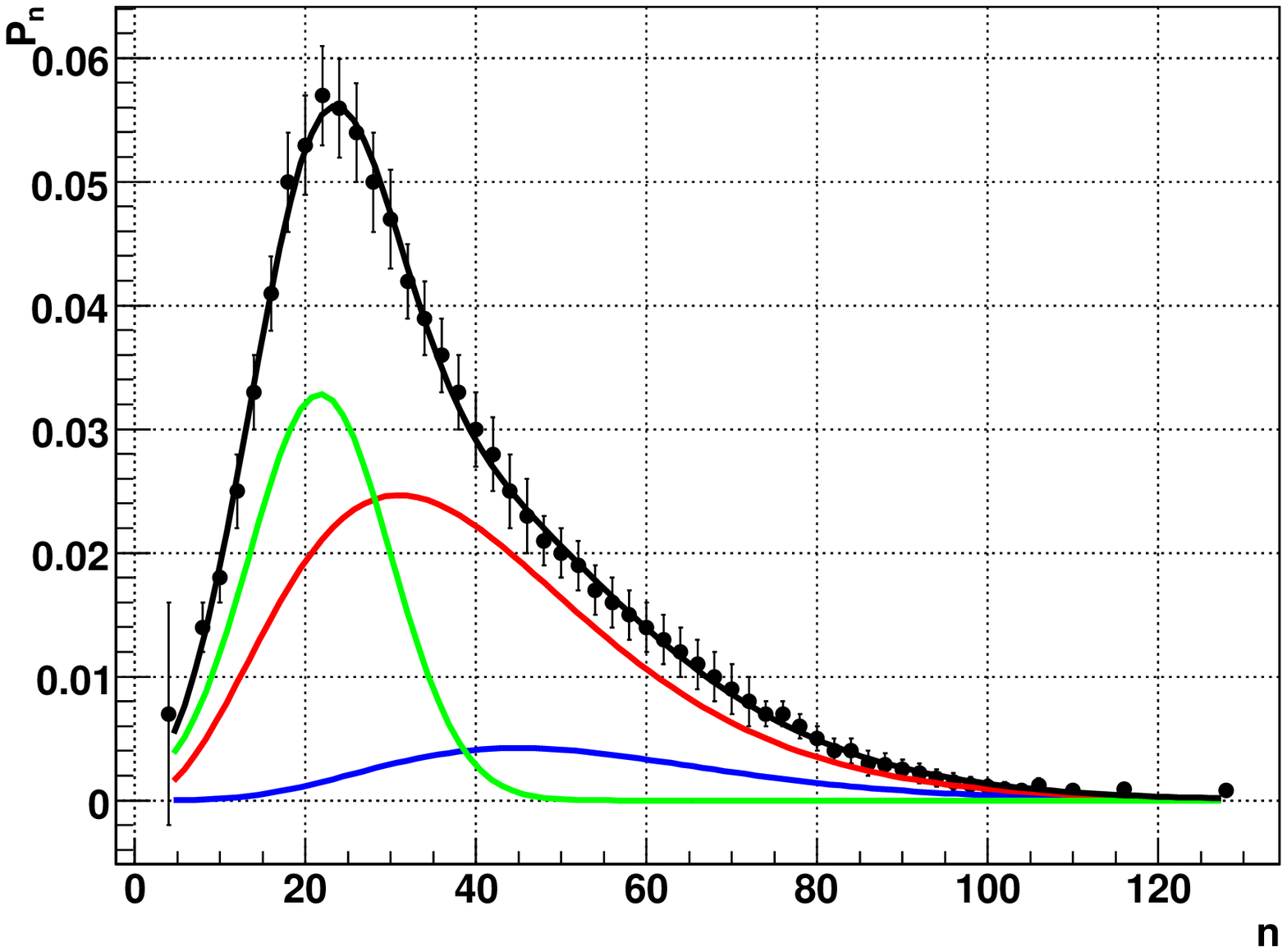}}
\caption{Charged  multiplicity distribution for proton-antiproton
scattering, $\sqrt{s}=900$~GeV~\cite{bib10}. Blue line -- negative
binomial distribution for three quark strings, red line --
negative binomial distribution for two quark strings, green line
-- Gaussian distribution for gluon string, black line is sum of
these distributions, $\chi^2/ndf=8/49$. }
\end{figure}
\begin{figure}[!h]
\centerline{
\includegraphics[scale=0.65]{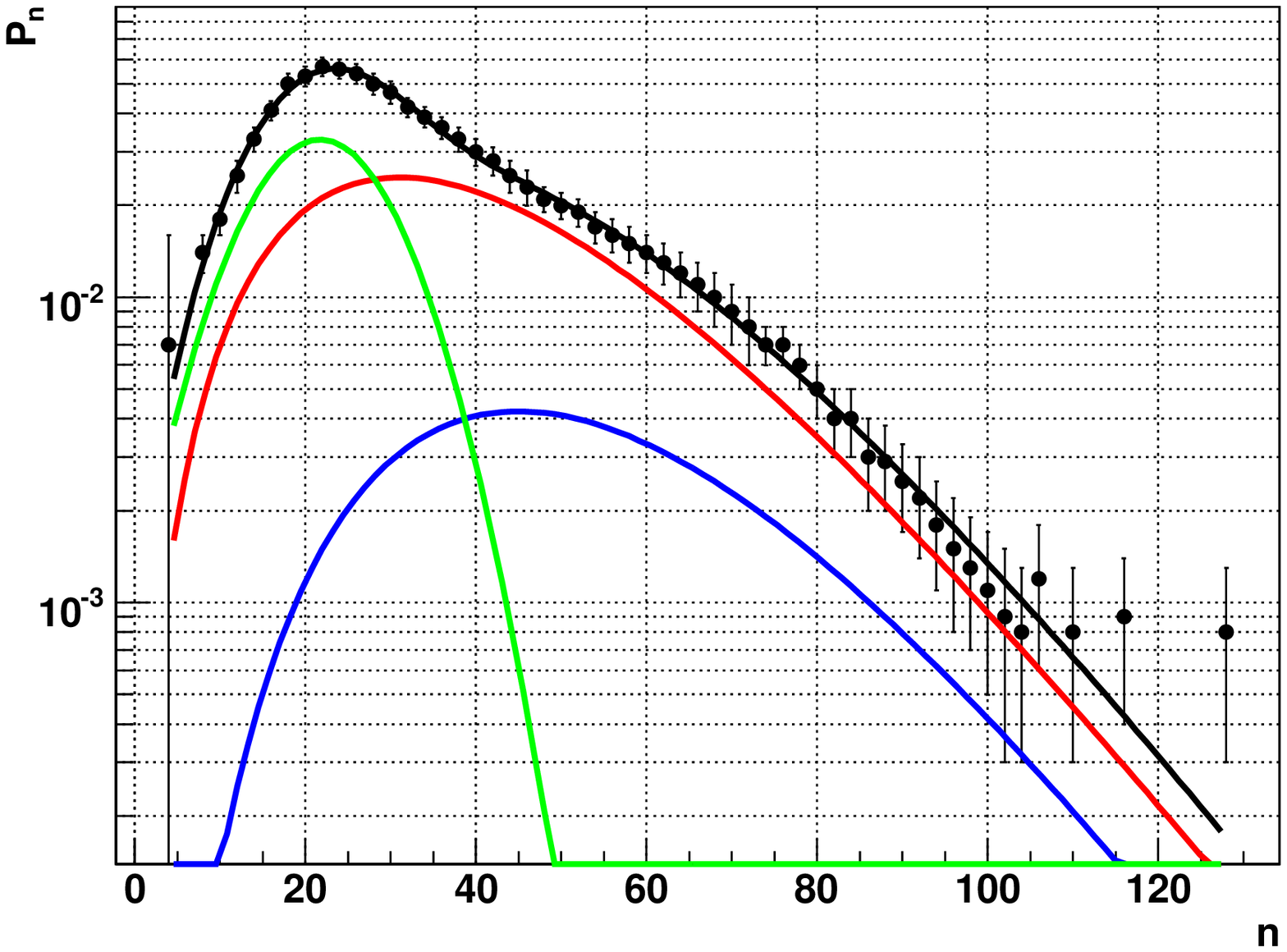}}
\caption{The same as for Fig. 19,  logarithmic scale.}
\end{figure}

\newpage
\begin{figure}[h]
\centerline{
\includegraphics[scale=0.65]{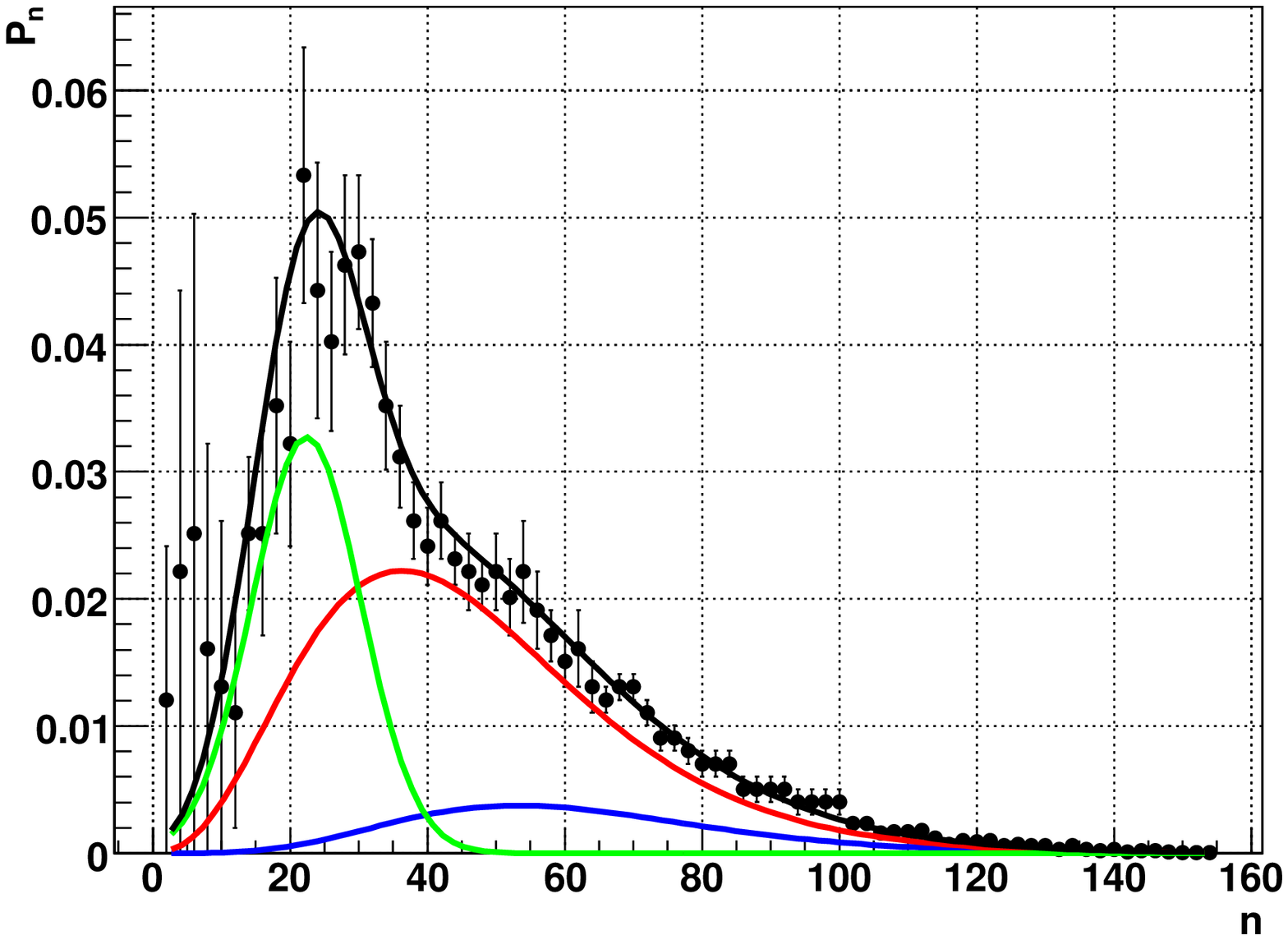}}
\caption{Charged  multiplicity distribution for proton-antiproton
scattering, $\sqrt{s}=1000$~GeV~\cite{bib11}. Blue line --
negative binomial distribution for three quark strings, red line
-- negative binomial distribution for two quark strings, green
line -- Gaussian distribution for gluon string, black line is sum
of these distributions, $\chi^2/ndf=53/73$. }
\end{figure}
\begin{figure}[!h]
\centerline{
\includegraphics[scale=0.65]{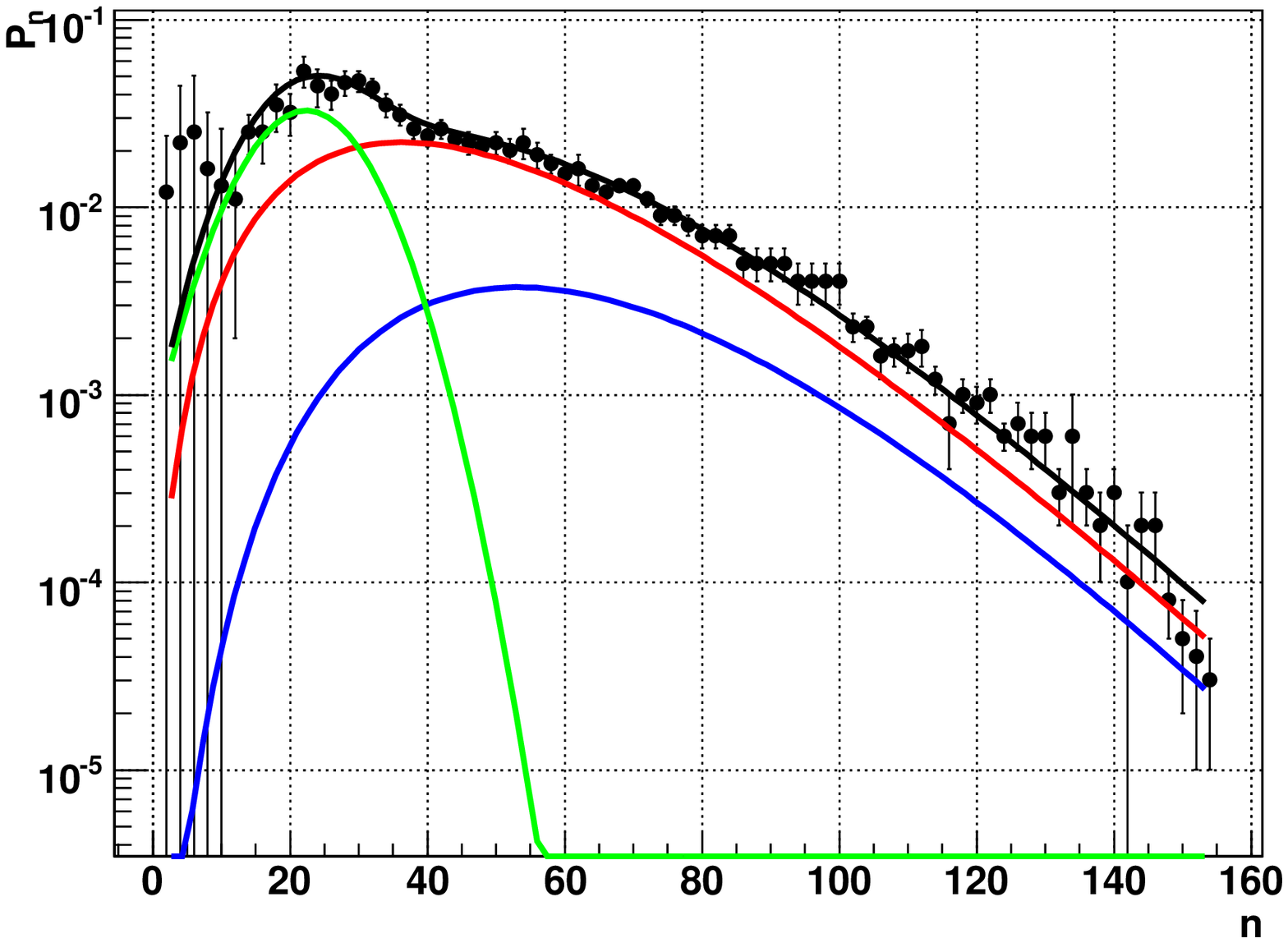}}
\caption{The same as for Fig. 21,  logarithmic scale.}
\end{figure}

\newpage
\begin{figure}[h]
\centerline{
\includegraphics[scale=0.65]{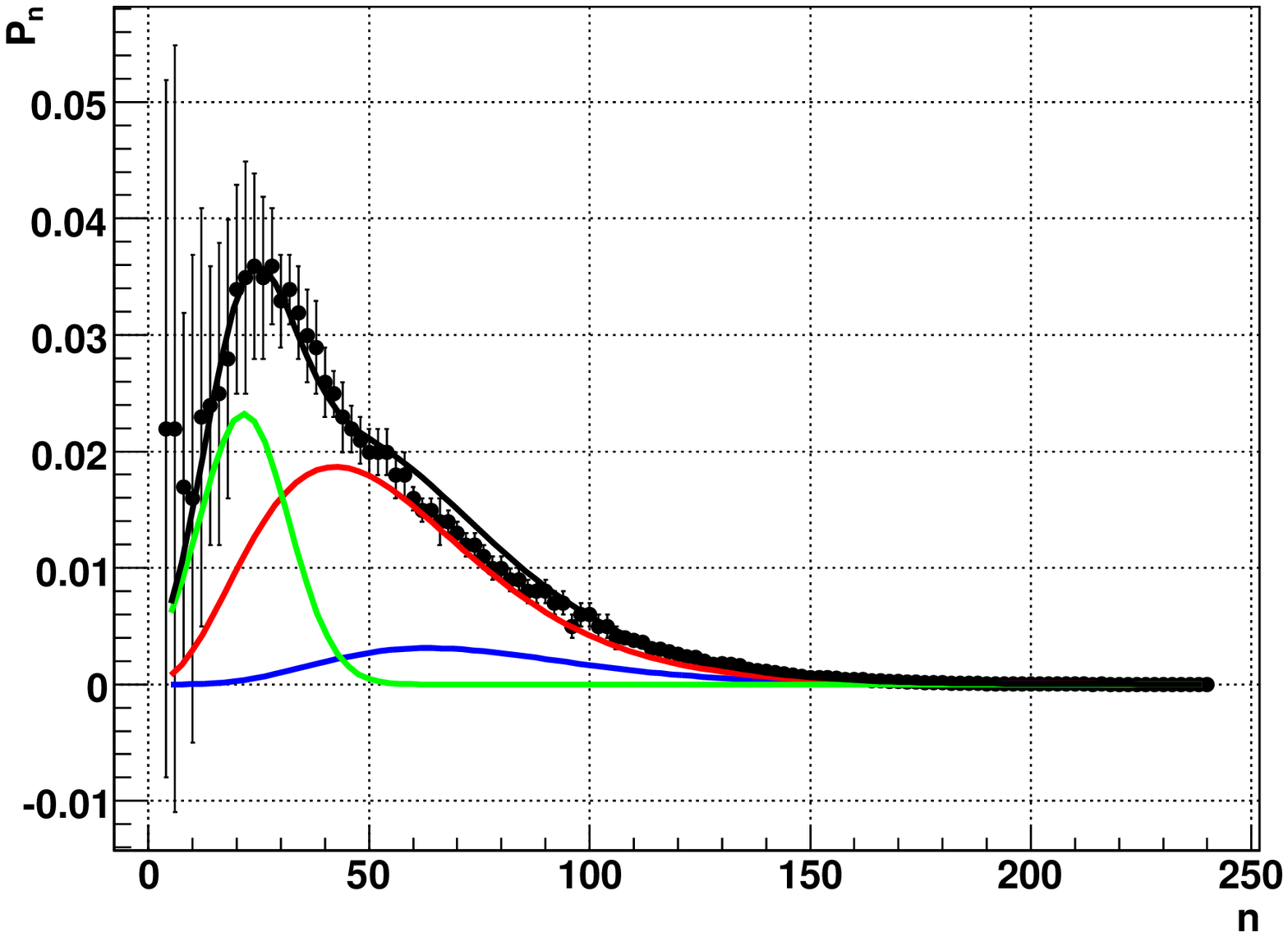}}
\caption{Charged  multiplicity distribution for proton-antiproton
scattering, $\sqrt{s}=1800$~GeV~\cite{bib11}. Blue line --
negative binomial distribution for three quark strings, red line
-- negative binomial distribution for two quark strings, green
line -- Gaussian distribution for gluon string, black line is sum
of these distributions, $\chi^2/ndf=120/115$. }
\end{figure}
\begin{figure}[!h]
\centerline{
\includegraphics[scale=0.65]{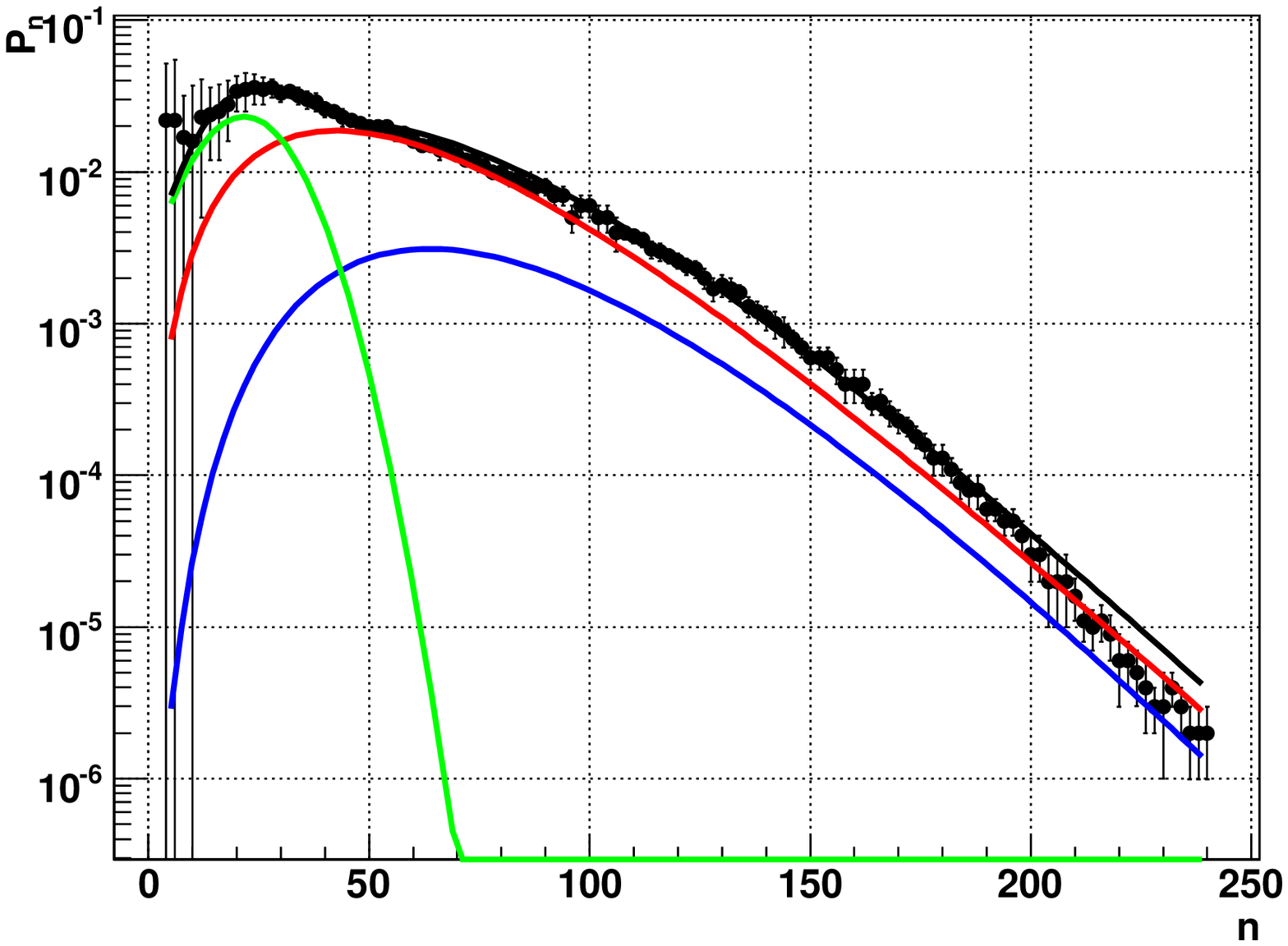}}
\caption{The same as for Fig. 23,  logarithmic scale.}
\end{figure}

\newpage
\begin{figure}[h]
\centerline{
\includegraphics[scale=0.65]{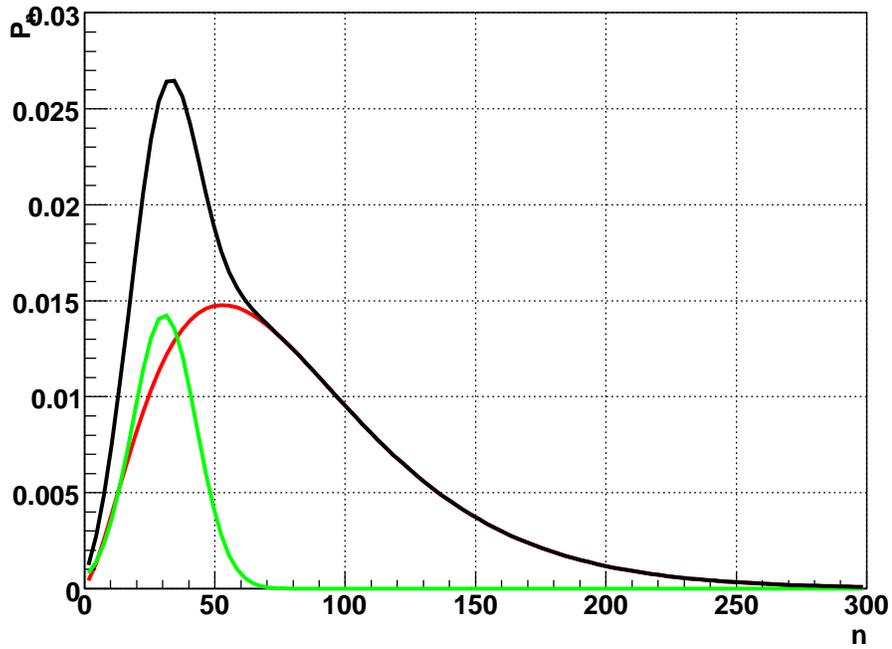}}
\caption{Charged  multiplicity distribution for proton-antiproton
scattering, $\sqrt{s}=14$~TeV. Red line -- negative binomial
distribution for two quark strings, green line -- Gaussian
distribution for gluon string, black line is sum of these
distributions.}
\end{figure}
\begin{figure}[!h]
\centerline{
\includegraphics[scale=0.65]{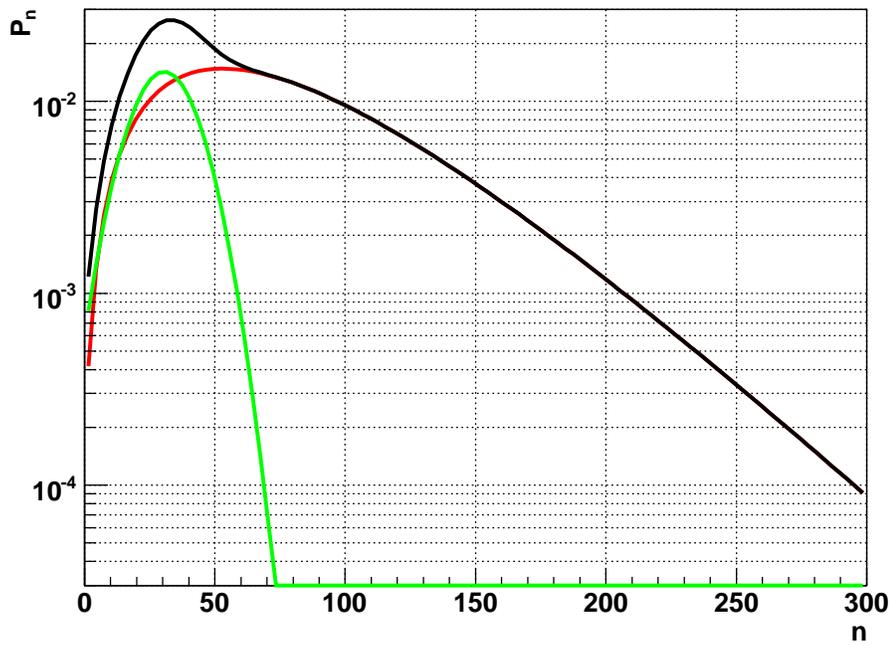}}
\caption{The same as for Fig. 25,  logarithmic scale.}
\end{figure}

\section{Conclusion}
The main results of this work are the following: we have
calculated the charged  multiplicity distribution and its mean
value $\langle n \rangle=71.57\pm4.37$, and total cross section
$\sigma_{tot}^{pp}= 101.30\pm6.65$~mb for energy
$\sqrt{s}=14$~TeV.

It was shown that multiplicity distributions are determined by
contributions of  inelastic processes of new type. These inelastic
processes completely differ from inelastic processes originating
from unitary cuts of pomeron and pomeron
branchings~\cite{bib12},~\cite{bib13}. These are processes of
hadrons production in gluon string and in two quark strings for
proton-proton scattering and processes of hadrons production in
gluon string and in two and three quark strings for
proton-antiproton scattering.

It is necessary to emphasize that inelastic processes for
proton-proton scattering differ from inelastic processes for
proton-antiproton scattering.

We are grateful to O.V.~Kancheli for useful discussions.

\end{document}